\documentclass[twocolumn,secnumarabic,amssymb, nobibnotes, aps, prm]{revtex4-1}
\usepackage{amsfonts}
\usepackage{graphicx}
\usepackage{epsfig}
\usepackage{epsf}
\usepackage{amssymb}
\usepackage{amsmath}
\usepackage{amsthm}
\usepackage{multirow}
\usepackage[usenames, dvipsnames,table,xcdraw]{xcolor}
\usepackage{mathrsfs}
\usepackage{bm}
\usepackage{dcolumn}
\usepackage{epstopdf}
\usepackage{bm,bbm}
\usepackage{cases,booktabs}
\usepackage[a4paper, pagebackref=false, colorlinks=true,
linkcolor=blue, citecolor=blue,
pdfauthor={ },
pdftitle={ },
pdfsubject={ },
pdfkeywords={ }]{hyperref}

\makeatletter
\newcommand\figcaption{\def\@captype{figure}\caption}
\newcommand\tabcaption{\def\@captype{table}\caption}
\makeatother

\begin{document}

\title{Determining the direction of a nanowire's flexural vibrations by micro-lens optical fiber interferometer}

\author{Chenghua Fu$^{1,2,3}$}
\author{Wen Deng$^{1,2,3}$}
\author{Lvkuan Zou$^{1,3}$}
\email[]{zoulvkuan@hmfl.ac.cn}
\author{Wanli Zhu$^{1,2,3}$}
\author{Ning Wang$^{1,3}$}
\author{Fei Xue$^{1,3}$}
\email[]{xuef@hmfl.ac.cn}

\affiliation{
$^{1}$Anhui Province Key Laboratory of Condensed Matter Physics at Extreme Conditions, High Magnetic Field Laboratory, Chinese Academy of Sciences, Hefei 230031, Anhui, People's Republic of China\\
$^{2}$University of Science and Technology of China, Hefei 230026, People's Republic of China\\
$^{3}$Collaborative Innovation Center of Advanced Microstructures, Nanjing University, Nanjing 210093, People's Republic of China}

\date{\today}

\begin{abstract}
Nanowires are perfect transducers for ultra-sensitive detections of force and mass. For small mass sensing, recent advances in detecting more properties than their masses with high sensitivity for absorbed particles onto a nanowire rely on identifications of two orthogonal flexural vibration modes of the nanowire. The directions of these orthogonal flexural vibrations with respect to measurement direction are crucial parameters of vibration modes. However, previous method, which determines a nanowire's vibration direction using thermal vibrations, requires simultaneously detecting of two orthogonal flexural vibrations of the nanowire with sufficient sensitivity. In this work, we propose and realize a method for the determination of directions of a nanowire's flexural vibrations by micro-lens optical fiber interferometer. Our method combines the light interference and light scattering of the nanowire. It does not require detecting a pair of degenerated orthogonal flexural vibration modes. Therefore, our method is expected to have wide usages in characterizing nanowire's vibrations and their applications.

\end{abstract}

\maketitle

\section{Introduction}
Since the invention of atomic force microscopy (AFM), cantilevered mechanical resonator has experienced a tremendous boost in both basic research and applications. Over the past decades, it has developed as a versatile and powerful platform to measure varieties of physical quantities related to mechanical vibration due to its small suspended mass, low stiffness and high frequency bandwidth \cite{Yamaguchi2017semist,Wang2017advsci}. Quantities to be measured are transferred to frequency shift of cantilever vibration. By monitoring displacement or vibration of cantilevers with electrical or optical method, one can derive quantity of interest with high sensitivity.

For cantilevers with sufficiently small damping coefficient, which is the case in most experiments, resonant frequency relies on both modal mass m and stiffness k of the cantilever: $\omega = \sqrt{k/m}$, where $\omega$ is angular resonant frequency. This equation provides two transparent routines to tune frequency by stiffness and by modal mass, which correspond to force detection and mass sensing, respectively. Dynamical cantilever magnetometry (DCM) is a typical practice by mainly tuning stiffness of cantilever \cite{Weber2012nanolett,Mehlin2015nanolett,Seo2017apl}. The recently developed method has shown advantages with at least three orders of magnitude enhancement compared with commercial superconducting quantum interference device(SQUID). For interdisciplinary studies, such as biophysics concerning molecule adsorption and nanoelectromechanical system(NEMS) \cite{Kosaka2014nnanotech}, cluster adsorption behavior could be given by resonant frequency shift tuned by effective modal mass. This simple relation is valid when added mass can be regarded as the change of effective mass of the whole cantilever, which works in small particles adsorbed on relatively large cantilever. However, it sets a limitation for small mass detection and could not meet the requirement for more and more versatile and precise mass sensing.

In 2010, Eduardo Gil-Santos and the collaborators proposed a novel method to determine both mass and position of adsorption particles a nanowires with the aid of investigation of two orthogonal vibration eigenmodes of a nanowire \cite{Santos2010nnanotech}. They found that mass change of adsorption particle could cause the two vibration modes to rotate with respect to optical measurement axis and observed dependence of frequency splitting on adsorption position. This pioneering work shed light on the unique role of multiple mechanical mode \cite{Sakuma2012apl,Vallabhaneni2013jva,Davidovikj2016nanolett} on precise measurement for physical quantities \cite{Zhang2013nanolett}. In addition to conventional force detection \cite{Garcia2012nnanotech,Doolin2014njp,Moser2013nnanotech,Lepinay2017nnanotech,Meier2016nanotech,Weber2016ncommun}, displacement measurement \cite{Ramos2013srep} and small mass sensing \cite{Chaste2012nnanotech,Malvar2013apl,Hanay2015nnanotech,Zhang2015actb,Olcum2015ncommun} including single molecule identification \cite{Hanay2012nnanotech}, analysis of multiple mechanical eigenmodes could also give critical information of vectorial detection \cite{Gloppe2013cleo,Gloppe2014nnanotech,Rossi2017nnanotech}, mechanical resonator cooling \cite{Ramos2012nanolett}, nonlinear coupling between two modes \cite{Cadeddu2016nanolett,Foster2016nanolett}, frequency fluctuations \cite{Sansa2016nnanotech} and energy decay \cite{Guttinger2017nnanotech}.

Owing to these promising attributes for practice and abundant physical insight, study of multiple mechanical eigenmodes of cantilevers has drawn great attention and research interest of the community. Among parameters of mechanical modes, vibration orientation, i.e., the angle with respect to optical axis or measurement direction, is one of significance, which affects subsequent analysis of the whole detection system. In previous studies, the angle is mostly simulated by theoretical models \cite{Kosmaca2017nanotech,Marghitu2012springer} like Ritz formulation \cite{Liew1993solidsstruct,Kang2004mechsci} with customized morphology utilized in the experiment. In 2011, Kiracofe et al. put forward a direct experimental determination scheme for the orientation of mechanical vibration modes using thermally driven vibrations \cite{Kiracofe2011nanotech}. They designed a nanowire cantilever with an elliptical cross section to artificially obtain frequency-splitted two nearly degenerated orthogonal eigenmodes of mechanical vibrations. By monitoring relative magnitude of resonant mode peaks in deflection spectrum density and taking the equipartition theorem into account, the angle $\theta$ between mode orientation and measurement direction is:
\begin{equation}
 \theta=tan^{-1}\sqrt{{\frac{\langle x^2_{1,obs}\rangle\omega^2_1}{\langle x^2_{2,obs}\rangle\omega^2_2}}} ,
\label{eq01}
\end{equation}
where $\langle x^2_{j,obs}\rangle$ and $\omega_j$ are observed mean squared deflection (MSD) and angular resonant frequency in the $j$th eigenmode($j = 1, 2$) respectively. It is assumed that directions of two vibration modes $\omega_j = \sqrt{k_j/m}$ are orthogonal. These assumptions may break down in certain systems, then determination of the angle will be a problem.

In this work, we propose a new experimental method to determine the direction of a nanowire's flexural vibrations by micro-lens optical fiber interferometer. Based on combined interference and scattering effect of light interacting with nanowires using thermally driven vibration, the mode angle can be reconstructed through the projections i.e., corresponding MSDs, of vibration eigenmodes to two orthogonal reference axes. Comparing with Kiracofe's scheme which relies on simultaneously detecting of two orthogonal flexural vibrations of the nanowire with sufficient sensitivity, this new method provides an independent and robust characterization for individual vibration mode and is expected to offer a new perspective for nanomechanical ultrasensitive detection.

\section{Proposal and theoretical analysis}

\begin{figure}[btp]
\centering
\includegraphics[width=8.5cm]{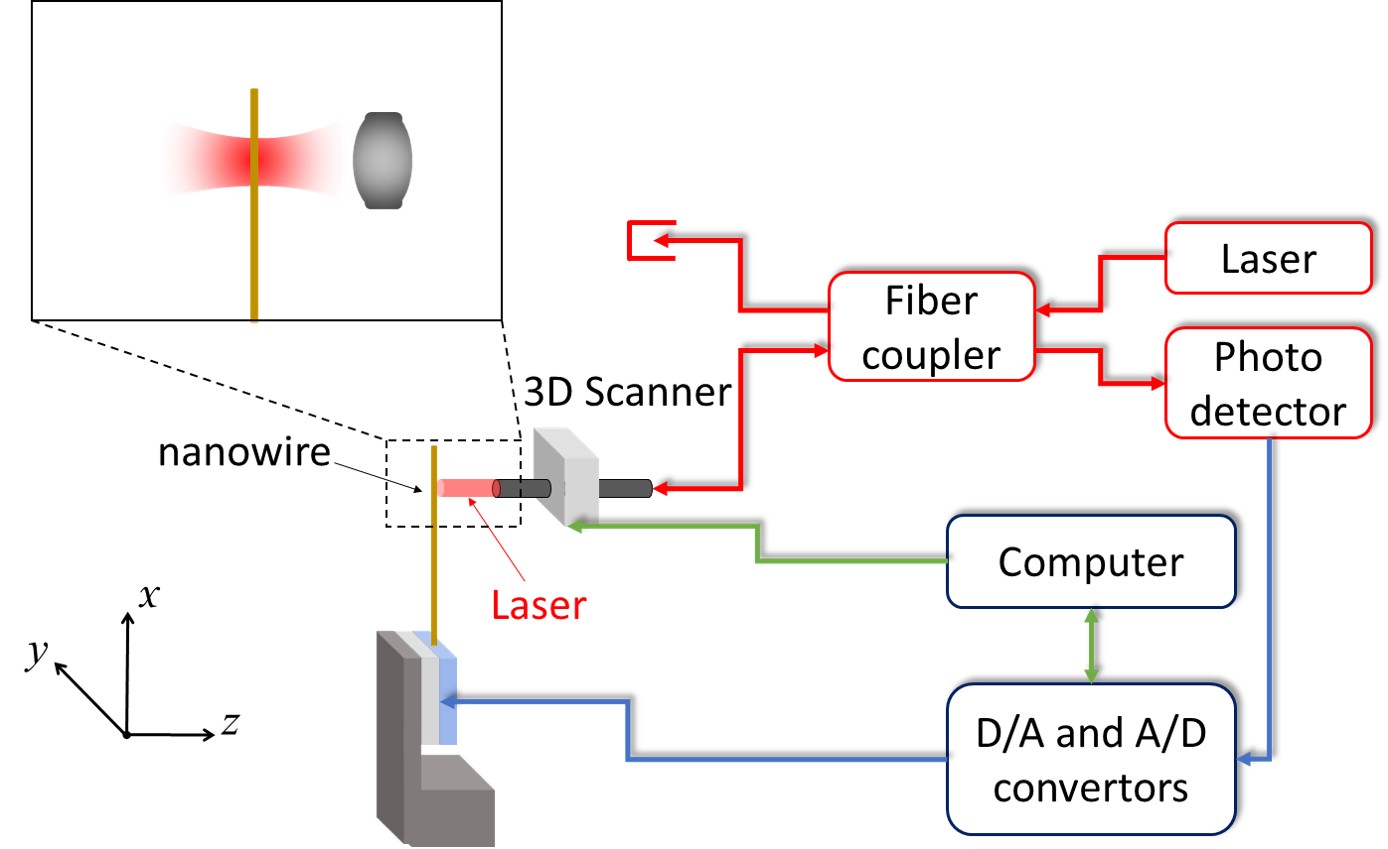}
\caption{(Color online) Schematic of micro-lens optical fiber interferometer. The inset shows zoom-in relationship between micro-lens and nanowire positioned at the center of outgoing Gaussian beam focal spot.}
\label{fig:setup}
\end{figure}

In our proposal, vibration eigenmode of nanowire was measured by self-made micro-lens optical fiber interferometer as shown in Fig. (\ref{fig:setup}), which consists of light emission/receiving part (represented by red frames and lines), detection core (including displacement scanner and nanowire holder) and electrical signal collection/processing part (connected by green and blue lines). The optical fiber coupler has a split ratio of 90:10 thus reducing output optical power to one tenth, a moderate degree that thermal effect could be neglected.

The probe head (detection core) where light interacts with nanowire is sealed in a vacuum chamber. A three dimensional (3D) scanner device with nanometer precision controlled by computer is crucial to implement the new proposal in this work. Nanowire is mounted on a delicate structure fixed with the entire framework. As the zoom-in part of Fig. (\ref{fig:setup}) shows, we locate the focal spot of exit Gaussian laser beam around the free end of nanowire to maximize the observed vibration amplitude. Since our nanowire cantilever can be applied with Euler-Bernoulli beam theory, for the lowest order pair of degenerated orthogonal flexural vibration modes, displacement or vibration amplitude scales monotonically with the distance from the fixed end.

As for electrical signal collection/processing part, multiple instruments including digital-to-analog and/or analog-to-digital converter, lock-in amplifier, low noise preamplifier, low-pass filter and data acquisition device along with the computer may be used to perform the experiment.

Nanowire cantilever used in our experiment is made from a commercial single crystal silicon tipless cantilever(NanoWorld) by focused ion beam (FIB) technique as Fig. \ref{fig:PSD}(b) shows.

The new method proposed in this work relies on coherent interplay of interference and scattering effect when light interacts with nanowire cantilever. The pure interference effect is solely sensitive to the distance change along optical axis. For the periodic sinusoidal dependence on distance, this effect is naturally suitable for measurement of displacement comparable with laser wavelength, approximately micrometer scale, hence it has been frequently utilized for a powerful tool as been reported by previous literatures. Here we notice the scattering effect when laser interacts with nanowire. For the case where size of nanowire and laser wavelength are comparable, as in this work, the mechanism can be well analyzed by Mie scattering theory.


\begin{figure}[ht]
\centering
\includegraphics[width=6.9cm]{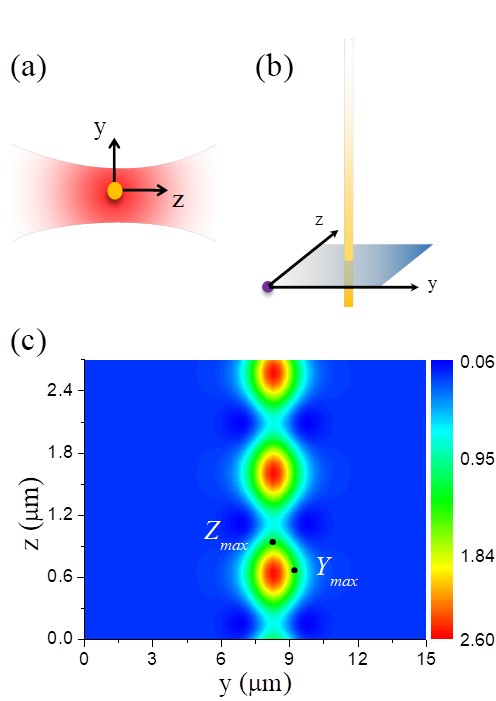}
\caption{(Color online) Determination of working points. (a) Top view of the nanowire positioning at the center of Gaussian laser beam focal spot. (b) Scanning region in y-z plane and the nanowire. The purple point indicates scanning start point. (c) Output voltage from photo detector v.s. the position of nanowire: determining working points. The color bar represents the output voltage of photo detector. These numerical values are fitting result of experimental data for scanning in y-z plane. Y$_{max}$ working point and Z$_{max}$ working point are shown in black dots.}
\label{fig:Gaussianbeam}
\end{figure}

Based on the analysis above, we can obtain a mixed contribution of the two effects when we scan the focal spot in the plane enclosed by y and z axis as shown in Fig. \ref{fig:Gaussianbeam}(b). The output voltage from photo detector is:
\begin{eqnarray}\label{eq02}
V_{out}(y,z) &=& C\mid A+G(y)\,e^{ i 4\pi\frac{z-z_0}{\lambda}}\mid ^2, \nonumber \\
G(y) &=& D + E\, e^{-\frac{(y-y_0)^2}{2\,w_0^2}},
\end{eqnarray}
where A and C are parameters related to the instrument, i.e., the micro-lens optical fiber interferometer, G(y) is a Gaussian shape interacting function determined by Mie scattering of the nanowire and laser, $\lambda$ is the wavelength of laser. We consider Taylor expansion for a sufficiently small vibration amplitude at a working point ($y,z$) in the y-z plane:
\begin{eqnarray}\label{eq03}
V_{out}(y+\delta y,z+\delta z)&\approx& V_{out}(y,z)+C_y \,\delta y+C_z \, \delta z, \nonumber\\
C_y &=& 2 C [A \, \cos(\frac{4\pi}{\lambda}z) + G(y)]\frac{\partial G(y)}{\partial y} , \nonumber \\
C_z &=& - 2C[ A G(y)\frac{4\pi}{\lambda} \sin(\frac{4\pi}{\lambda}z)],
\end{eqnarray}
where partial derivatives $C_y$ and $C_z$ denote conversion coefficients corresponding to change of output voltage caused by small displacement $\delta_y$ along y direction and small displacement $\delta_z $ along z direction, respectively. In Eq. (\ref{eq03}), we have let $z_0 =0$ for convenience. These two coefficients are crucial to the analysis presented below for mode angle determination.

We define `Y$_{max}$ working point' for the one with the largest absolute value $\mid C_y\mid$ among all the possible local minimum and maximum working points which gives the highest sensitivity for displacement or vibration projection in y direction and does not respond to z component, or, equivalently, only visible for scattering effect. Therefore, at the Y$_{max}$ working point, we have
\begin{eqnarray}\label{eq04}
C_z &=&  0,  \nonumber \\
\frac{\partial C_y}{\partial y} &=& 0,  \nonumber  \\
\frac{\partial C_y}{\partial z} &=& 0, \nonumber  \\
det(Y) &=& \begin{vmatrix}
             \frac{\partial^2 C_y}{\partial y^2} & \frac{\partial^2 C_y}{\partial y \partial z} \\
             \frac{\partial^2 C_y}{\partial z \partial y} & \frac{\partial^2 C_y}{\partial z^2} \\
           \end{vmatrix} \neq 0.
\end{eqnarray}
Solutions to Eq. (\ref{eq04}) give all possible local minimum and maximum critical points in the y-z plane with the exclusion of saddle points by restricting matrix Y to be positive definite or negative definite as Hessian matrix second derivative test requires.

Correspondingly, the `Z$_{max}$ working points' is defined by
\begin{eqnarray}\label{eq05}
C_y &=&  0,  \nonumber \\
\frac{\partial C_z}{\partial y} &=& 0,  \nonumber  \\
\frac{\partial C_z}{\partial z} &=& 0, \nonumber  \\
det(Z) &=& \begin{vmatrix}
             \frac{\partial^2 C_z}{\partial y^2} & \frac{\partial^2 C_z}{\partial y \partial z} \\
             \frac{\partial^2 C_z}{\partial z \partial y} & \frac{\partial^2 C_z}{\partial z^2} \\
           \end{vmatrix} \neq 0,
\end{eqnarray}
along with conditions that matrix Z is positive definite or negative definite and $\mid C_z\mid$ is the largest. Similarly, `Z$_{max}$ working points' is the best point to measure interference effect.

In an experiment, we can find the two typical working points by following procedures. (i) Locate the free end of nanowire to the center of Gaussian beam by three dimensional (3D) high precision piezo positioning system, as shown in Fig. \ref{fig:Gaussianbeam}(a). (ii) Scan the focal spot in y-z plane with respect to the nanowire's position and record the output voltage from the photo detector. (iii) Fit experimental data with Eq. (\ref{eq02}), two dimensional profile of output voltage could be plotted which gives the value of $\mid C_y \mid _{max}$ and $\mid C_z \mid _{max}$ along with corresponding coordinates of Y$_{max}$ working points and Z$_{max}$ working points as labeled in Fig. \ref{fig:Gaussianbeam}(c).

With the coefficients known, the other essential ingredient of our proposal is the measurement of power spectrum densities $S_{VV}$ at the two typical working points from thermal vibrations of nanowire. Considering the free end vibration x(t) of nanowire, its power spectral density has the form of
\begin{equation}\label{eq06}
  S_{xx}(f)=\frac{C_{x}}{(f^2_n-f^2)^2+(\frac{f_n f}{Q_n})^2}.
\end{equation}
Here $C_x$ is related to the amplitude of vibrations, $f_n$ and $Q_n$ are natural (resonant) frequency and quality factor of the observed eigenmode, respectively. The conversion coefficient plays an important role for connecting the popular used power spectral density (PSD) $S_{VV}$ with $S_{xx}$. For an arbitrary working point $P$ with conversion coefficient $C_P^2 =  C_y^2 + C_z^2$, namely, the magnitude of gradient of $V_{out}$ at that point, the relation reads
\begin{equation}\label{eq07}
  S_{VV}(f)= C_P^2 S_{xx}(f).
\end{equation}

At Y$_{max}$ working points and Z$_{max}$ working points, $C_P$ will be coincident with $C_y$ and $C_z$. We calculate the integral of both sides of Eq. (\ref{eq07}) over the full frequency range and conclude the mean squared deflection(MSD) $\langle x^2 _{obs}\rangle$
\begin{equation}\label{eq08}
  \int^{\infty}_0 S_{VV}(f)df= C_P ^2 \int^{\infty}_0 S_{xx}(f)df= C_P ^2 \langle x^2 _{obs}\rangle,
\end{equation}
where $\langle x^2 _{obs}\rangle$ is observed MSD of eigenmode, or, equivalently, the projection of eigenmode vibration $\langle x^2 \rangle$ along the working points's gradient of $C_P$.

Suppose the orientation of one mechanical eigenmode deviates from the measurement direction, i.e., the z axis, with an angle of $\theta$ as in Fig. \ref{fig:PSD}(b). The observed MSD at Y$_{max}$ working points is then
\begin{equation}\label{eq09}
  \langle x^2 _{y,obs}\rangle=\langle x^2\rangle \sin^2\theta =\frac{\int^\infty _0 S^y _{VV}(f)df}{\mid C_y^2\mid_{max}}\equiv\frac{I_y}{ \mid C_y ^2 \mid_{max}},
\end{equation}
where we denote the integral measured at Y$_{max}$ working points $\int^\infty_0 S^y_{VV}(f)df$ as $I_y$. Correspondingly, result at Z$_{max}$ working points gives
\begin{equation}\label{eq10}
  \langle x^2 _{z,obs}\rangle=\langle x^2\rangle \cos^2\theta =\frac{\int^\infty _0 S^z _{VV}(f)df}{ \mid C_z ^2 \mid_{max}}\equiv\frac{I_z}{ \mid C_z ^2 \mid_{max}}.
\end{equation}
By solving simultaneous equations of Eq. (\ref{eq09}) and Eq. (\ref{eq10}), we eventually obtain eigenmode angle $\theta$
\begin{equation}\label{eq11}
  \theta=tan^{-1}\sqrt{\frac{ \mid C_z ^2\mid _{max}}{ \mid C_y^2\mid _{max}}\frac{I_y}{I_z}},
\end{equation}
along with actual MSD
\begin{equation}\label{eq12}
  \langle x^2 \rangle =\frac{I_y}{\mid C_y^2\mid_{max}} + \frac{I_z}{\mid C_z^2\mid_{max}}.
\end{equation}

\section{Experiment Result}
\begin{figure}[b]
\centering
\includegraphics[width=8.8cm]{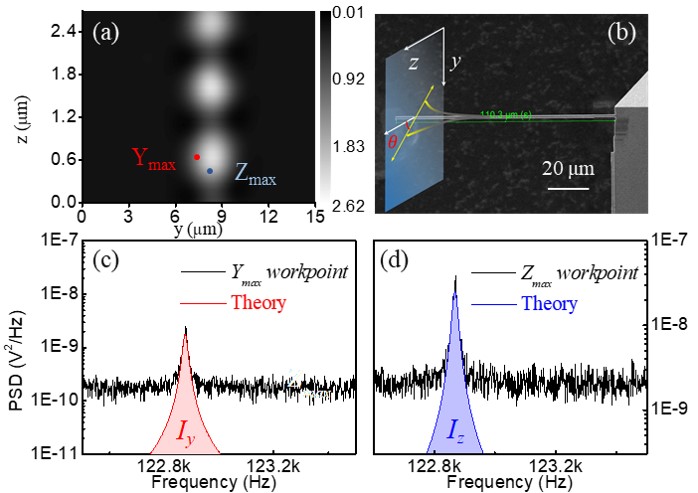}
\caption{(a) Dependence of output voltage from photo detector in unit of voltage on laser spot position in the y-z plane. (b) Scanning electron microscopy image of the nanowire cantilever used in this work and the illustration of direction of flexural vibration mode. (c),(d) Measured PSDs at Y$_{max}$ working point and Z$_{max}$ working point, respectively. Red solid line and blue solid line represent fitting results at Y$_{max}$ working point and Z$_{max}$ working point respectively while red and blue region are integrals corresponding to I$_y$ and I$_z$.}
\label{fig:PSD}
\end{figure}

Experiment measured scanning output voltage from photo detector is plotted in Fig. \ref{fig:PSD}(a) as a gray scale contour plot. The data is consistent with theoretical analysis and surface fitting result in Fig. \ref{fig:Gaussianbeam}(c) with Eq. (\ref{eq02}) gives positions of Y$_{max}$ working points and Z$_{max}$ working points. The corresponding conversion coefficients are extracted:
\begin{eqnarray}\label{eq13}
  \mid C_y\mid _{max} &=& 27.672 \,\, \text{V/$\mu$m}, \nonumber\\
  \mid C_z\mid _{max} &=& 89.577 \,\, \text{V/$\mu$m}.
\end{eqnarray}

Fig. \ref{fig:PSD}(b) is the scanning electron microscopy (SEM) image of nanowire. In order to adjust the resonant frequency of eigenmode within proper range and emphasize the influence of morphological asymmetry to dynamic behavior of nanowire resonator, we made some modification at the connection part of nanowire cantilever with holder plate, including a step like morphology (terrace) fine-processing by FIB and deposition of platinum thin film by electron beam evaporation. The free end vibration is illustrated as the yellow bending part of nanowire and the yellow arrow indicates vibration direction of mechanical eigenmode in y-z plane. Mode angle with respect to measurement direction (z axis), $\theta$, can be clearly identified in this setup.

The experiment is conducted at room temperature and high vacuum(10$^{-2}$ mbar) conditions so that the Q factor of eigenmode can be high enough for efficient identification of resonant peaks in PSD measurement. We position focal spot to Y$_{max}$ working points and Z$_{max}$ working points to acquire thermally induced vibration PSDs and investigate the eigenmode with resonant frequency of about 123kHz. Figures \ref{fig:PSD}(c) and \ref{fig:PSD}(d) show PSD data at Y$_{max}$ working points and Z$_{max}$ working points, respectively. Taking fitted conversion coefficients and integrals (I$_y$,I$_z$) into Eq. (\ref{eq11}), flexural vibration angle $\theta$ can be calculated:
\begin{eqnarray}\label{eq14}
  \theta &=& tan^{-1}\sqrt{\frac{\mid C_z^2\mid_{max}}{\mid C_y^2\mid_{max}}\frac{I_y}{I_z}} \nonumber \\
   &=& tan^{-1}\sqrt{\frac{(89.577V/\mu m)^2}{(27.672V/\mu m)^2}\times \frac{5.5\times 10^{-8}V^2}{8.3\times10^{-7}V^2}} \nonumber \\
   &=& 39.8^{\circ},
\end{eqnarray}
which coincides with simulated result by COMSOL software. Eq. (\ref{eq12}) gives thermally driven vibration amplitude $\sqrt{\langle x^2 \rangle}\approx 13.2$ pm.

The validity and universality of this method is guaranteed by high stability and detection sensitivity of our micro-lens optical fiber interferometer. Owing to large numerical aperture (NA) of micro-lens and some procedures to suppress instrument noise, the spatial resolution, which is determined by the laser spot size, is in the order of micrometers. The displacement sensitivity for nanowire's flexural vibrations' measurements, characterized by noise floor in PSD measurements, is about 0.5 pm/rtHz, which is among the best of similar optical interferometer with simple setups \cite{Santos2010nnanotech,Tamayo2012nanotech,Tamayo2013csr,Ramos2013srep,Pigeau2015ncommun,Lee2017nanoscale}. With increasing complexity and/or adopting new nanowire materials, the displacement sensitivity may be further improved to quantum regime, typically at fm/rtHz or higher \cite{Poot2012physrep,Doolin2014njp,Weber2016ncommun,Barg2017apb,Tsaturyan2017nnanotech}. Taking resonant frequency 122.9 kHz and Q factor 6000 of the eigenmode investigated in this work at room temperature and 10$^{-2}$ mbar for example, the smallest detectable vibration amplitude of nanowires is about 3 pm. The micro-lens fiber optical interferometer system has been carefully adjusted to ensure optimal measurement.

\section{Conclusion}
In summary, we propose a new method to unambiguously determine eigenmode vibration angle based on combining interference with scattering effect of laser interacting with nanowire cantilevers. Parameters of single eigenmode could be directly retrieved from experiments regardless of the relative configuration of two modes and/or nonlinear coupling between them. This result is expected to pave the way for future advances of ultrasensitive detection and dynamics study with multi eigenmodes of mechanical resonators.

\section*{Acknowledgments}
This work was supported by the National Natural Science Foundation of China, Grants No. 11374305, No. 11604338, and No. 11704386; the Technological Development Grant of Hefei Science Center of Chinese Academy of Sciences, Grants 2014TDG-HSC001; the National Key Research and Development Program of China, Grants 2017YFA0303201; F. X. also thanks the support of  Recruitment Program for Young Professionals.

\newpage

\end{document}